\documentclass[twocolumn,showpacs,amsmath,amssymb]{revtex4}
\usepackage{graphicx}
\usepackage{amsfonts}
\usepackage{dcolumn}
\usepackage{bm}

\usepackage{CJK}

\begin{document}
\title{Explosive Phase Transition in a Majority-Vote Model with Inertia}

\author{Hanshuang Chen$^{1}$}\email{chenhshf@ahu.edu.cn}

\author{Chuansheng Shen$^{2,3}$}

\author{Haifeng Zhang$^4$}

\author{Guofeng Li$^1$}

\author{Zhonghuai Hou$^{5}$}\email{hzhlj@ustc.edu.cn}

\author{J\"urgen Kurths$^{2,6}$}\email{Juergen.Kurths@pik-potsdam.de}

\affiliation{$^{1}$School of Physics and Materials Science, Anhui University, Hefei, 230601, China \\
$^2$Department of Physics, Humboldt University, 12489 Berlin, Germany \\
$^3$Department of Physics, Anqing Normal
University, Anqing, 246011, China\\
$^4$School of Mathematical Science, Anhui University, Hefei, 230601, China \\
$^5$Hefei National Laboratory for Physical Sciences at Microscales
\& Department of Chemical Physics, University of
 Science and Technology of China, Hefei, 230026, China\\
$^6$Potsdam Institute for Climate Impact Research (PIK), 14473
Potsdam, Germany }

\date{\today}

\begin{abstract}
We generalize the original majority-vote model by incorporating an
inertia into the microscopic dynamics of the spin flipping, where
the spin-flip probability of any individual depends not only on the
states of its neighbors, but also on its own state. Surprisingly,
the order-disorder phase transition is changed from a usual
continuous type to a discontinuous or an explosive one when the
inertia is above an appropriate level. A central feature of such an
explosive transition is a strong hysteresis behavior as noise
intensity goes forward and backward. Within the hysteresis region, a
disordered phase and two symmetric ordered phases are coexisting and
transition rates between these phases are numerically calculated by
a rare-event sampling method. A mean-field theory is developed to
analytically reveal the property of this phase transition.

\end{abstract}
 \maketitle

Phase transitions in ensembles of complex networked systems have
been a subject of intense research in statistical physics and many
other disciplines \cite{RMP08001275}. These results are of
fundamental importance for understanding various dynamical processes
in real world, such as percolation
\cite{PhysRevLett.85.4626,PRL00005468}, epidemic spreading
\cite{RevModPhys.87.925}, synchronization
\cite{PRP08000093,PRP2016}, and collective phenomena in social
networks \cite{RMP09000591}.

Recently, explosive or discontinuous transitions in complex networks
have received growing attention since the discovery of an abrupt
percolation transition in random networks
\cite{Science323.1453,PhysRevLett.103.255701} and scale-free
networks \cite{PhysRevLett.103.168701,PhysRevLett.103.135702}. Later
studies affirmed that this transition is actually continuous but
with an unusual finite size scaling
\cite{PhysRevLett.105.255701,PhysRevLett.106.225701,Science333.322},
yet many related models show truly discontinuous and anomalous
transitions (cf. \cite{NatPhys2015.11} for a recent review).
Striking different from continuous phase transitions, in an
explosive transition an infinitesimal increase of the control
parameter can give rise to a considerable macroscopic effect.
Subsequently, an explosive phenomenon was found in the dynamics of
cascading failures in interdependent networks
\cite{Nature2010,PhysRevLett.105.048701,PhysRevLett.107.195701}, in
contrast to the second-order continuous phase transition found in
isolated networks. More recently, such explosive phase transitions
have been reported in various systems, such as explosive
synchronization due to a positive correlation between the degrees of
nodes and the natural frequencies of the oscillators
\cite{PhysRevLett.106.128701,PhysRevLett.108.168702,PhysRevLett.110.218701}
or an adaptive mechanism \cite{PhysRevLett.114.038701},
discontinuous percolation transition due to an inducing effect
\cite{NatComm2013}, spontaneous recovery \cite{NatPhys2014}, and
explosive epidemic outbreak due to cooperative coinfections of
multiple diseases \cite{EPL2013,NatPhys2015,PNAS2015}.

In this paper we report an explosive order-disorder phase transition
in a generalized majority-vote (MV) model by incorporating the
effect of individuals' inertia (called \emph{inertial MV model}).
The MV model is one of the simplest nonequilibrium generalizations
of the Ising model that displays a continuous order-disorder phase
transition at a critical value of noise \cite{JSP1992}. It has been
extensively studied in the context of complex networks, including
random graphs \cite{PhysRevE.71.016123,PA2008}, small world networks
\cite{PhysRevE.67.026104,IJMPC2007,PA2015}, and scale-free networks
\cite{IJMPC2006(1),IJMPC2006(2)}. However, the continuous nature of
the order-disorder phase transition is not affected by the topology
of the underlying networks \cite{PhysRevE.91.022816}. In our model,
we have included a substantial change to make it more realistic,
namely the state update of each node depends not only on the states
of its neighboring nodes, but also on its own state. In fact, in a
social or biological context individuals have a tendency for beliefs
to endure once formed. In a recent experimental study, behavioral
inertia was found to be essential for collective turning of starling
flocks \cite{NatPhys2014(2)}. We refer this modification as
\emph{inertial effect}. Surprisingly, we find that as the level of
the inertia increases, the nature of the order-disorder phase
transition is changed from a continuous second-order transition to a
discontinuous, or an explosive, first-order one. For the latter
case, a clear hysteresis region appears in which the order and
disordered phases are coexisting. In particular, a relevant
phenomenon of inertia-induced first-order synchronization transition
was found in a second-order Kuramoto model
\cite{PhysRevLett.78.2104,PhysRevE.89.022123}. A counterintuitive
``slower is faster" effect of the inertia on ordering dynamics of
the voter model was reported in a recent work
\cite{PhysRevLett.101.018701}.

We first describe the original MV model defined on underlying
networks. Each node is assigned to a binary spin variable ${\sigma
_i} \in \{ + 1, - 1\}$ $(i = 1, \ldots ,N)$. In each step, a node
$i$ is randomly chosen and tends to align with the local
neighborhood majority but with a noise parameter $f$ giving the
probability of misalignment. In this way, the single spin-flip
probability from $\sigma_i$ to $-\sigma_i$ can be written as
\begin{eqnarray}
w(\sigma_i)=\frac{1}{2}\left[{1-(1-2f)\sigma_i S\left( {\Theta
_i}\right)}\right] \label{eq1}
\end{eqnarray}
with
\begin{eqnarray}
\Theta_i=\sum\limits_{j=1}^N a_{ij}\sigma_j \label{eq2}
\end{eqnarray}
where $S(x)=sgn(x)$ if $x\neq0$ and $S(0)=0$. The elements of the
adjacency matrix of the underlying network are defined as $a_{ij}=1$
if nodes $i$ and $j$ are connected and $a_{ij}=0$ otherwise.

In the original MV model, the state update of each node depends
exclusively on the states of its neighboring nodes, regardless of
its own state. Here, we incorporate the inertial effect into the
original model by replacing Eq.(2) with
\begin{eqnarray}
\Theta_i =(1-\theta)\sum\limits_{j=1}^N a_{ij}\sigma _j/k_i + \theta
\sigma_i\label{eq3}
\end{eqnarray}
where $k_i=\sum\nolimits_{j = 1}^N a_{ij}$ is the degree of node
$i$, and $\theta \in [0,0.5]$ is a parameter controlling the weight
of the inertia. The larger the value of $\theta$ is, the larger the
inertia of the system is. For $\theta=0$, we recover to the original
MV model where no inertia exists. For $\theta=0.5$, our model is
dominated by the inertia other than the random spin flip with the
probability $f$. In this case, there is no spontaneous magnetization
to appear. If $\theta=0.5$ and $f=0$, the spins are frozen into the
initial configuration. We should note that the generalization of the
inertial MV model from two states to multiple states is
straightforward, which is discussed in the Supplementary Material
\cite{SM1}.

\begin{figure}
\centerline{\includegraphics*[width=1.0\columnwidth]{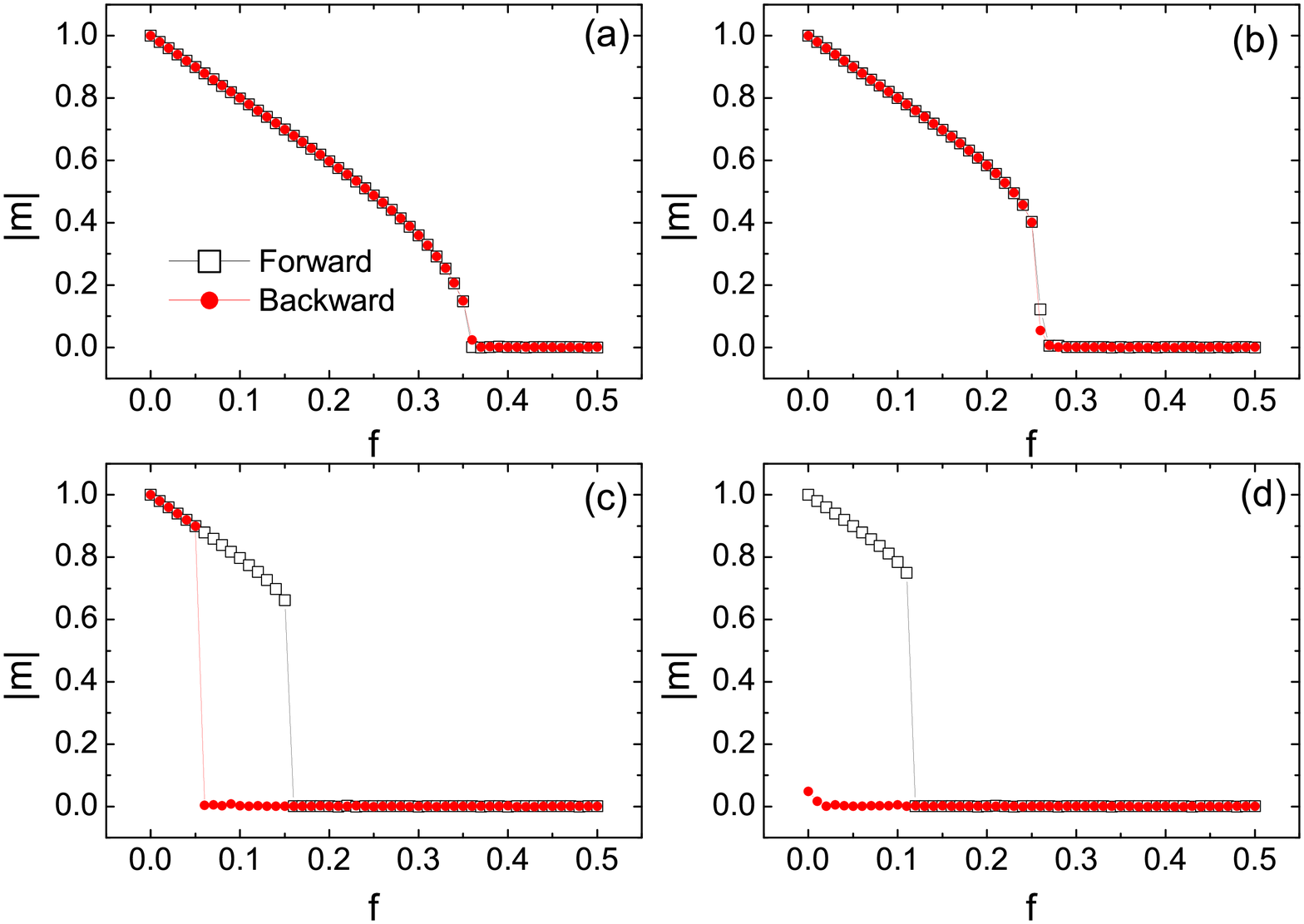}}
\caption{(color online). The absolute magnetization $|m|$ as a
function of noise intensity $f$ on ER random networks with different
inertia parameter $\theta=0$ (a), $\theta=0.2$ (b), $\theta=0.3$
(c), and $\theta=0.35$ (d). The symbols squares and circles
correspond to forward and backward simulations, respectively. The
network parameters are $N=10,000$ and $\left\langle k
\right\rangle=20$. \label{fig1}}
\end{figure}

The phase behavior of the system can be characterized by the average
magnetization per node, $m=\sum\nolimits_{i=1}^N \sigma_i/N$. $m=0$
for the disordered phase and $m\neq0$ for the ordered phase. By
Monte Carlo (MC) simulations, Fig.\ref{fig1} shows the absolute
value of $m$ as a function of $f$ for several different values of
$\theta$ on Erd\"os-R\'enyi (ER) random networks (ER-RN) with the
size $N=10,000$ and the average degree $\left\langle k \right\rangle
=20$. The simulation results are obtained by performing forward and
backward simulations, respectively. The former is done by
calculating the stationary value of $m$ as $f$ increases from 0 to
0.5 in steps of 0.01, and using the final configuration of the last
simulation run as the initial condition of the next run, while the
latter is performed by decreasing $f$ from 0.5 to 0 with the same
step. For $\theta=0$, the results on the forward and backward
simulations coincide, implying that the order-disorder transition is
a continuous second-order phase transition that is the main feature
of the original MV model. For $\theta=0.2$, although the transition
becomes sharper and the transition point shifts to a smaller value
of $f$, the forward and backward simulations still coincide.
Strikingly, for $\theta=0.3$, one can see that as $f$ increases,
$|m|$ abruptly jumps from nonzero to zero at $f=f_{c_F}$, which
shows that a sharp transition takes place for the order-disorder
transition (Fig.\ref{fig1}(c)). On the other hand, the curve
corresponding to the backward simulations also shows a sharp
transition from the disordered phase to the order phase at
$f=f_{c_B}$. These two sharp transitions occur at different values
of $f$, leading to a clear hysteresis loop with respect to the
dependence of $|m|$ on $f$. Such a feature indicates that a
discontinuous first-order order-disorder transition arises due to
the effect of inertia. Further increasing $\theta$ to $\theta=0.35$,
$f_{c_F}$ shifts to a smaller value and $f_{c_B}$ decreases to zero,
but the nature of a discontinuous phase transition is still present.

\begin{figure}
\centerline{\includegraphics*[width=1.0\columnwidth]{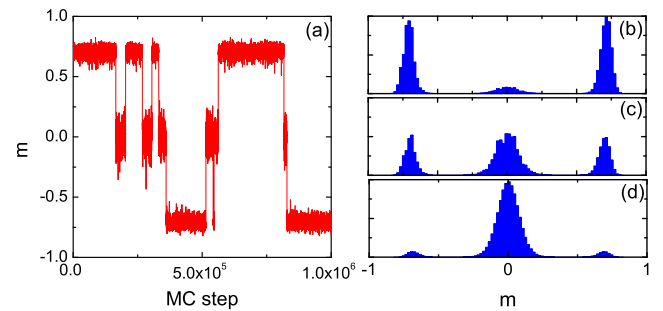}}
\caption{(color online). (a) Time series of magnetization $m$ show
the transition events between ordered and disordered phases within
the hysteresis region. (b-d) show the PDF of $m$ for $f=0.136$ (b),
$f=0.138$ (c), and $f=0.14$ (d). The other parameters are
$\theta=0.3$, $N=500$, and $\left\langle k \right\rangle=20$.
\label{fig2}}
\end{figure}

\begin{figure}
\centerline{\includegraphics*[width=1.0\columnwidth]{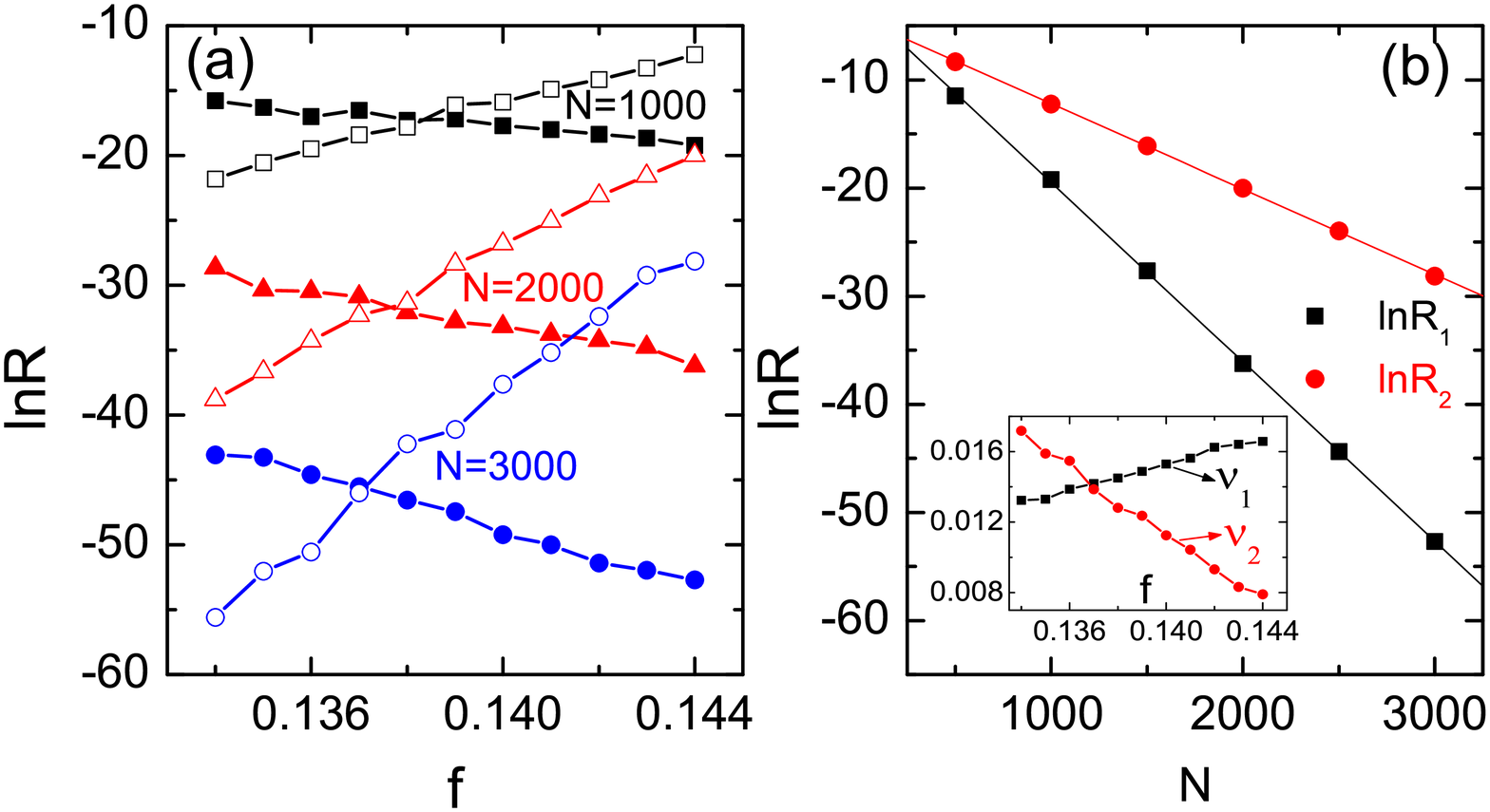}}
\caption{(color online). (a) The logarithm of transition rates $\ln
R$ as a function of $f$ for different $N$. Solid symbols correspond
to the transition rate from disordered state to ordered state, and
empty symbols to the transition rate from ordered state to
disordered state. (b) $\ln R$ as a function of $N$ for $f=0.144$.
The lines indicate the linear fitting $\ln R_{1(2)} \sim -
\nu_{1(2)}N$. The inset shows the fitting exponents $\nu_{1(2)}$ as
a function of $f$. \label{fig3}}
\end{figure}

Within the hysteresis region, we observe phase flips between the
ordered phase and the disordered one for a rather small network size
$N$, as shown in Fig.\ref{fig2}(a) by a long time series of $m$ in a
ER network of $N=500$. We show in Fig.\ref{fig2}(b-d) the
probability density function (PDF) of $m$ for three distinct $f$
chosen from the hysteresis region. On the one hand, all of them are
multimodal distributions with a peak at $m=0$ and two other peaks
symmetrically located at both sides of it. On the other hand, with
the increase of $f$ the peak at $m=0$ becomes higher implying that
the disorder phase becomes more stable. To calculate the transition
rates between the ordered and disordered phases, a long-time
simulation is necessary. However, for a larger network size the
transition rates are extremely low and brute-force simulation is
prohibitively expensive. To overcome this difficulty, we have used a
recently developed rare-event simulation method, the forward flux
sampling (FFS) \cite{PRL05018104,JPH09463102}. In Fig.\ref{fig3}(a),
we show the transition rates $R_1$ from disordered to ordered phases
and the inverse transition rate $R_2$ as a function of $f$ for
several different $N$. $R_1$ is a deceasing function of $f$ and
$R_2$ is an increasing function of $f$. The intersection point of
both curves determines the location at which the ordered and the
disordered phase are equally stable. As $N$ increases, the
intersection point slightly shifts to a smaller value. In
Fig.\ref{fig3}(b), we show the transition rates as a function of $N$
at $f=0.144$. Obviously, both $R_1$ and $R_2$ decrease exponentially
with $N$, $R_{1(2)} \sim \exp \left(- \nu_{1(2)}N\right)$ with the
exponents $\nu_{1(2)}$, implying that the disordered and ordered
phases are coexisting in the thermodynamic limit. In the inset of
Fig.\ref{fig3}(b), we give the fitting exponents $\nu_{1(2)}$ as a
function of $f$, and they clearly exhibit the different variation
trends with $f$.

In the following, we will present a mean-field theory to understand
the simulation results. We first define $m_k$ as the average
magnetization of a node of degree $k$, and $\tilde m$ as the average
magnetization of a randomly chosen nearest-neighbor node. For
uncorrelated networks, the probability that a randomly chosen
nearest-neighbor node has degree $k$ is $k P(k)/\left\langle k
\right\rangle$, where $P(k)$ is the degree distribution defined as
the probability that a node chosen at random has degree $k$ and
$\left\langle k \right\rangle$ is the average degree
\cite{RMP08001275}. Thus, $m_k$ and $\tilde m$ satisfy the following
relation
\begin{eqnarray}
\tilde m= \sum\limits_k k P(k) m_k / \left\langle k \right\rangle.
\label{eq4}
\end{eqnarray}

For an up-spin node $i$ of degree $k$, the probability that its
local field is positive can be written as the cumulative binomial
distribution,
\begin{eqnarray}
P_>^+ = \sum\limits_{n =\left\lceil {n_k^ + } \right\rceil }^k
{\left( {1 - \frac{1}{2}{\delta _{n,n_k^+}}} \right)}
C_k^n{p_\uparrow ^n}{p_\downarrow^{k - n}}.\label{eq5}
\end{eqnarray}
Here, $p_{\uparrow(\downarrow)}=(1 \pm \tilde m)/2$ is the
probability that a randomly chosen nearest-neighbor node has $+1$
($-1$) state, $\left\lceil \cdot \right\rceil$ is the ceiling
function, $\delta$ is the Kronecker symbol, $C_k^n =k!/[n!(k - n)!]$
are the binomial coefficients, and $n_k^+=(1 - 2\theta)k/[2(1 -
\theta )]$ is the number of up-spin neighbors of node $i$ satisfying
$\Theta _i=0$. Similarly, we can write the probability that the
local field of a down-spin node of degree $k$ is positive as,
\begin{eqnarray}
P_>^- = \sum\limits_{n = \left\lceil {n_k^ - } \right\rceil }^k
{\left( {1 - \frac{1}{2}{\delta _{n,n_k^-}}} \right)}
C_k^n{p_\uparrow ^n}{p_\downarrow^{k - n}},\label{eq6}
\end{eqnarray}
where $n_k^-=k-n_k^+=k/[2(1 - \theta )]$.

Furthermore, the spin-flip probability $\omega _k^ +$ of an up-spin
node of degree $k$ can be expressed as the sum of two parts.
\begin{eqnarray}
\omega_k^+ = fP_>^+  + (1-f)(1-P_>^+),\label{eq7}
\end{eqnarray}
where the first part is that the local field of the node is positive
and the minority rule is applied, and the other one is that the
local field of the node is negative and the majority rule is
applied. Likewise, we can write the spin-flip probability of a
down-spin node of degree $k$ as,
\begin{eqnarray}
\omega_k^- = f(1-P_>^-) + (1-f)P_>^-\label{eq8}
\end{eqnarray}

\begin{figure}
\centerline{\includegraphics*[width=1.0\columnwidth]{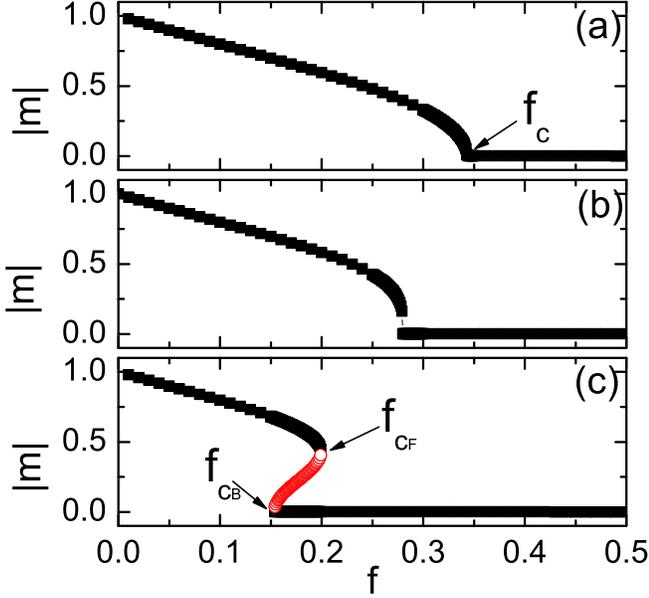}}
\caption{(color online). Theoretical result of $|m|$ as a function
of $f$ for three typical values of $\theta$: (a) $\theta=0.15$, (b)
$\theta=0.23$, and (c) $\theta=0.3$. In Fig.4(c), circles within the
hysteresis region indicate the unstable solution. \label{fig4}}
\end{figure}

\begin{figure}
\centerline{\includegraphics*[width=1.0\columnwidth]{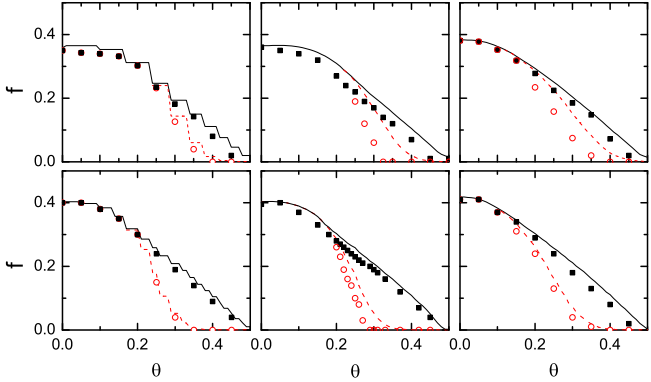}}
\caption{(color online). Phase diagram in the $\theta-f$ plane for
three types of networks with two different average degree:
$\left\langle k \right\rangle=20$ (top panels) and $\left\langle k
\right\rangle=40$ (bottom panels). For left to right: Rd-RN, ER-RN,
and BA-SFN. Lines and symbols correspond to the theoretical and
simulation results, respectively. $f_{c_F}$ are indicated by solid
lines and squares, and $f_{c_B}$ by dashed lines and circles. The
sizes of all the networks are the same: $N=10,000$. \label{fig5}}
\end{figure}

Thus, the rate equations for $m_k$ are
\begin{eqnarray}
\dot m_k =  - \left( {\frac{1+m_k}{2}} \right)\omega_k^+  + \left(
{\frac{1 -m_k}{2}} \right)\omega_k^- \label{eq9}
\end{eqnarray}
In the steady state $\dot m_k=0$, we have
\begin{eqnarray}
m_k = \frac{\omega _k^ -- \omega _k^ +  }{\omega _k^ + +\label{eq1}
\omega _k^ - }\label{eq10}
\end{eqnarray}
Inserting Eq.(\ref{eq10}) into Eq.(\ref{eq4}), we get a
self-consistent equation of $\tilde m$,
\begin{eqnarray}
\tilde m =\Psi(\tilde m),\label{eq11}
\end{eqnarray}
with
\begin{eqnarray}
\Psi(\tilde m)= \sum\limits_k {\frac{{kP(k)}}{{\left\langle k
\right\rangle }}\frac{{\omega _k^ - -\omega _k^ +}}{{\omega _k^ +  +
\omega _k^ - }}}. \nonumber
\end{eqnarray}
Since $P_>^+ + P_>^-=1$ and $\omega_k^ + =\omega _k^-$ at $\tilde m
=0$, one can easily check that $\tilde m =0$ is always a stationary
solution of Eq.(\ref{eq11}). This solution corresponds to a
disordered phase. The other possible solutions can be obtained by
numerically iterating Eq.(\ref{eq11}). Once $\tilde m$ is found, we
can immediately calculate $m_k$ by Eq.(\ref{eq10}) and the average
magnetization per node by $m=\sum\nolimits_k {P(k)m_k}$.

By a detailed numerical calculation for Eq.(\ref{eq11}) on ER-RN
with the Poisson degree distribution $P(k)=\left\langle k
\right\rangle ^k e^{ - \left\langle k \right\rangle }/k!$ and the
average degree $\left\langle k \right\rangle=20$, we find that the
critical value of $\theta$ is $\theta_c=0.23$. In Fig.\ref{fig4}, we
show the theoretical results on $|m|$ as a function of $f$ for three
typical values of $\theta=0.15$, 0.23, and 0.3. For
$\theta<\theta_c$, the order-disordered phase transition is of
continuous second-order type. For $\theta>\theta_c$, the phase
transition is of discontinuous first-order type and a clear
hysteresis loop appears. At $\theta=\theta_c$, the order parameter
$|m|$ has still a jump at $f=f_c$ but no hysteresis loop exists.

At the critical noises, $f_{c_F}$ and $f_{c_B}$, the
susceptibilities $\tilde \chi  = \partial \tilde m /  \partial f$
are diverging. According to Eq.(\ref{eq11}), the condition is
equivalent to
\begin{eqnarray}
\frac{{\partial \Psi }}{{\partial \tilde m}} &=& (1 -
2f)\sum\limits_k {\frac{{k P(k)}}{{\left\langle k \right\rangle }}}
[ {(\omega_k^ -  + \omega_k^ + )^{ - 1}}(\frac{{\partial P_ > ^ +
}}{{\partial \tilde m}} + \frac{{\partial P_ > ^ - }}{{\partial
\tilde m}}) \nonumber\\ &+& (\omega_k^ - - \omega_k^ + ){(\omega_k^
-  + \omega_k^ + )^{ - 2}}(\frac{{\partial P_ > ^ + }}{{\partial
\tilde m}} - \frac{{\partial P_ > ^ - }}{{\partial \tilde m}})] = 1
\label{eq12}
\end{eqnarray}
Here, $\partial P_>^\pm  / \partial \tilde m$ can be derived from
Eq.(\ref{eq5}) and Eq.(\ref{eq6})
\begin{eqnarray}
 \frac{{\partial P_ > ^ \pm }}{{\partial \tilde m}} = \left( {1 -
\frac{1}{2}{\delta _{n_k^ \pm ,\left\lceil {n_k^ \pm } \right\rceil
}}} \right)\mathbb{P}\left( {\left\lceil {n_k^ \pm } \right\rceil
};k \right) \nonumber \\ +  \frac{1}{2}{\delta _{n_k^ \pm
,\left\lceil {n_k^ \pm } \right\rceil }}\mathbb{P}\left(
{\left\lceil {n_k^ \pm } \right\rceil + 1} ;k \right),\label{eq13}
\end{eqnarray}
where the function $\mathbb{P}(n;k)$ is defined as
\begin{eqnarray}
\mathbb{P}(n;k)=\frac{1}{2}k C_{k-1}^{n-1}p_\uparrow
^{n-1}p_\downarrow ^{k-n}\label{eq14}
\end{eqnarray}
For any given $\theta$, $f_{c_F}$ and $f_{c_B}$ are determined by
numerically solving Eqs.(\ref{eq11}-\ref{eq12}). In fact, $f_{c_B}$
can be obtained more conveniently, since $f_{c_B}$ corresponds to
the point at which the trivial solution $\tilde m=0$ loses its
stability. Therefore, $f_{c_B}$ is determined solely by
Eq.(\ref{eq12}). At $\tilde m=0$, Eq.(\ref{eq12}) can be reduced to
\begin{eqnarray}
{\left. {\frac{{\partial \Psi }}{{\partial \tilde m}}}
\right|_{\tilde m = 0}} = (1 - 2{f_{c_B}}){\sum\limits_k {\frac{{k
P(k)}}{{\left\langle k \right\rangle }}\left[
{\frac{{\frac{{\partial P_ > ^ + }}{{\partial \tilde m}} +
\frac{{\partial P_ > ^ - }}{{\partial \tilde m}}}}{{\omega_k^ -  +
\omega_k^ + }}} \right]} _{\tilde m = 0}}=1\label{eq15}
\end{eqnarray}

In Fig.\ref{fig5} we plot the phase diagram in the $\theta-f$ plane
for three types of networks (from left to right: random
degree-regular networks (Rd-RN), ER-RN, and Barab\'asi-Albert
scale-free networks (BA-SFN)) with two different average degrees:
$\left\langle k \right\rangle=20$ (top panels) and $\left\langle k
\right\rangle =40$ (bottom panels). The lines and symbols indicate
the theoretical and simulation results, respectively. For Rd-RN,
each node has the same degree $k$, which is a typical representation
of degree homogeneous networks. For BA-SFN, its degree distribution
follows a power-law function with the exponent $-3$, which is
typical for degree heterogeneous networks. Clearly, there is no
essential difference in the phase diagrams for different network
types and average degree. The phase diagram is divided into three
regions by $f_{c_F}$ and $f_{c_B}$. In the region below $f_{c_B}$,
the system is ordered. In the region above $f_{c_F}$, the system is
disordered. Between $f_{c_F}$ and $f_{c_B}$, the region is of
hysteresis with a disordered phase and two ordered phases of up-down
symmetry. As expected, for networks with a larger average degree the
mean-field theory provides a better prediction for the simulation
results. Although there exists obvious differences for a smaller
network connectivity, the theory and simulations are qualitatively
consistent.

In conclusion, we have investigated the order-disorder phase
transition in a MV model with inertia, where the inertia is
introduced into the state-updating dynamics of nodes by considering
the state of each node itself besides the states of its neighboring
nodes. We mainly find that in contrast to a continuous second-order
phase transition in the original MV model, the inertial MV model
undergoes a discontinuous first-order phase transition when the
inertia is large enough. In the hysteresis region of the first-order
phase transition, a disordered phase and two symmetric ordered
phases are coexisting. The transition rates between the disordered
and ordered phases have been calculated by FFS sampling. A
mean-field theory provides an analytical understanding for this
interesting phenomenon. Since behavioral inertia is an essential
characteristic of human being and animal groups, our work may shed a
novel understanding of transition phenomena from disorder to order,
like the emergence of consensus and decision-making
\cite{PhysRevLett.94.178701,PhysRevLett.116.038701}, as well as the
spontaneous formation of a common language/culture
\cite{RMP09000591,PhysRevLett.85.3536}. Finally, we expect further
investigations of inertial effect in other dynamical systems.

\begin{acknowledgments}
This work was supported by National Science Foundation of China
(Grants No. 11205002, 61473001, 11475003, 21473165), the Key
Scientific Research Fund of Anhui Provincial Education Department
(Grant No. KJ2016A015) and ``211" Project of Anhui University.
\end{acknowledgments}

%

\end{document}